\documentclass[11pt,a4paper]{article}

\usepackage[T1]{fontenc}
\usepackage[utf8]{inputenc}
\usepackage{lmodern}
\usepackage{amsmath,amssymb,amsthm}
\usepackage[margin=1in]{geometry}
\usepackage{graphicx}
\usepackage{natbib}
\usepackage{hyperref}
\IfFileExists{mathtools.sty}{\usepackage{mathtools}}{}
\IfFileExists{microtype.sty}{\usepackage{microtype}}{}
\IfFileExists{booktabs.sty}{\usepackage{booktabs}}{%
  \newcommand{\toprule}{\hline}%
  \newcommand{\midrule}{\hline}%
  \newcommand{\bottomrule}{\hline}%
  \providecommand{\addlinespace}[1][0pt]{}%
}
\IfFileExists{todonotes.sty}{\usepackage[disable]{todonotes}}{%
  \newcommand{\todo}[1]{}%
}

\newtheorem{proposition}{Proposition}
\newtheorem{definition}{Definition}
\theoremstyle{remark}
\newtheorem*{remark}{Remark}

\newcommand{\stratG}{\ensuremath{\pi_{\mathrm{G}}}}
\newcommand{\stratH}{\ensuremath{\pi_{\mathrm{H}}}}
\newcommand{\stratC}{\ensuremath{\pi_{\mathrm{C}}}}

\title{\textbf{Beyond the briscola advantage:\\
a Monte Carlo dominance test for deterministic strategies\\
in two-player Briscola Game}}

\author{%
  Piero Giacomelli\\
  \small Tenax S.p.A.\\
  \small \href{mailto:pgiacome@gmail.com}{\texttt{pgiacome@gmail.com}}%
}

\date{\today}

\begin{document}

\maketitle

\begin{abstract}
Briscola is a traditional Italian trick-taking card game whose simplest
form is played by two players. Popular folklore credits victory almost
entirely to the player who is dealt more cards of the trump suit (the
so-called \emph{briscola}), so that the game would be a near-deterministic
function of the deal. We test this folklore against a pre-registered
alternative, namely that two deterministic rule-based refinements of the
naive greedy policy --- a briscola-hoarding policy $\stratH$ and a
public-information counter policy $\stratC$ --- dominate the greedy
baseline $\stratG$ irrespective of trump luck. To this end we run a
round-robin Monte Carlo tournament of $10^{6}$ simulated games across the
nine ordered pairings of $(\stratG,\stratH,\stratC)$, retaining
approximately $1.08\times 10^{5}$ non-tied games per pairing, and we
analyse the resulting outcomes through Wilson confidence intervals, a
Bonferroni-corrected pairwise binomial test, and a logistic regression of
the game outcome on the strategy pair and on the signed briscola-count
imbalance, so as to quantify the relative contribution of strategy and
trump luck. Two findings run against the folklore in opposite
directions. First, the folk theorem fails in its deterministic form: the
player holding the briscola majority wins only $62.95\%$ of non-tied
games, with $95\%$ Wilson confidence interval $[62.83\%,\,63.06\%]$, so
that briscola luck is a strong but far from decisive predictor. Second,
the pre-registered dominance hypothesis is itself rejected, and in the
opposite direction to the one we had anticipated: the naive greedy policy
$\stratG$ strictly dominates both $\stratH$ and $\stratC$ in every
ordered pairing and at every Bonferroni-adjusted level, with odds-ratio
estimates $\mathrm{OR}_{\stratH/\stratG}=0.853$ $[0.844,\,0.862]$ and
$\mathrm{OR}_{\stratC/\stratG}=0.740$ $[0.732,\,0.747]$ when the policy
plays the first seat. Holding the strategy pair fixed, each additional
briscola held over the game multiplies the odds of winning by $1.22$
$[1.21,\,1.22]$. Strategy and trump luck are therefore both highly
significant and essentially orthogonal, with the strategy effect
substantially larger than the briscola-count effect. We close with a
reproducibility appendix that makes the simulation, the random seed and
the analysis script fully deterministic.
\end{abstract}

\textbf{Keywords:} applied Monte Carlo; trick-taking games; Briscola; strategy
dominance; logistic regression; reproducibility.

\section{Introduction}\label{sec:intro}

\begin{figure}[h]
\centering
\includegraphics{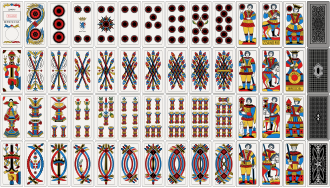}
\caption{Complete briscola card deck from Brescia region\cite{wikipediaBriscola}. }
\label{fig:deck}
\end{figure}

Briscola is a well-known Italian trick-taking card game, played with a
$40$-card regional deck and a single trump suit --- the
\emph{briscola} --- that is fixed at the start of the game by exposing
the last card of the stock. Although the game is essentially unknown
outside Italy, within Italy it is a genuine cross-generational pastime:
its rules are few and its pace is brisk, so that it serves both as a
common pedagogical entry point into trick-taking play for children and
as a stock social activity for older players in community centres and
family gatherings. From a formal point of view, Briscola is a game of
imperfect information and sequential strategic decisions, which makes
it a natural object of game-theoretic analysis. Its simplest form is
the two-player game on which the present paper concentrates; the game
also admits a four-player partnership variant played between two teams
of two, and a five-player variant in which one of the players is
secretly drawn to side with one team, so that the alliance structure
itself becomes hidden information during play. Each Italian region uses
its own deck artwork, which leaves the underlying rules unchanged but
visibly distinguishes, for instance, a Brescian deck from a Neapolitan
or a Piacentine one (see Figure~\ref{fig:deck}).

The linguistic origins of the word ``Briscola'' are themselves a
long-standing subject of debate among ludologists and linguists. The
most widely accepted theory derives the name from the French
\emph{brisque}, literally a ``galloon'', that is, the gold or silver
braid used on military uniforms to mark rank \citep{barganews2023}, and
is consistent with the popular tradition according to which the game
was a favourite pastime of French soldiers and Dutch sailors during
the late sixteenth and early seventeenth centuries. Older soldiers were
themselves nicknamed ``galloons'' (\emph{brisques}), and because the
game was so closely associated with these veterans, the term eventually
shifted from the soldiers to the game itself \citep{barganews2023}. A
competing strand of scholarship instead points to a Dutch or
Scandinavian origin, and emphasises the game's popularity among the
sailors who travelled to the Dutch East Indies \citep{barganews2023}.
In the Italian linguistic record the word first appears in the early
nineteenth century, and the earliest documented literary occurrence is
in an 1847 poem by Gioacchino Belli \citep{fenomenologia2020}.

Beyond its name, Briscola is generally regarded as a descendant of
\emph{Brusquembille}, an eighteenth-century French trick-taking game,
and is closely related to \emph{Bazzica}, known in English as
\emph{Bezique} \citep{wikipedia_briscola}. Interestingly, while the
name Briscola itself likely came from the French \emph{brisque},
Giampaolo Dossena has observed that the French name for Bazzica
(\emph{b\'esique}) may instead have Italian roots, which points to a
non-trivial cross-pollination of card-playing traditions across the
Alps \citep{fenomenologia2020}. Despite this rich cultural footprint,
Briscola has received comparatively little academic attention. The few
existing studies that engage with the game from a formal perspective
focus on the application of advanced artificial-intelligence
techniques, specifically Monte Carlo Tree Search and Deep
Reinforcement Learning, to the two-player
\citep{gazda2021game,singhdeep} and five-player
\citep{villa2012algoritmi} versions of the game. A formal
game-theoretic definition of Briscola and a systematic investigation of
its possible winning strategies, by contrast, have to our knowledge
never been carried out.

The present paper addresses this gap by investigating, on a large-scale
Monte Carlo basis, the folk claim widely shared among players that,
across the twenty tricks of a two-player game, whoever is dealt more
cards of the briscola suit \emph{necessarily} wins, so that the identity
of the winner would be essentially determined by the random deal rather
than by strategic decisions during play. The claim corresponds to a
sharp dichotomy between two stylised attitudes: the policy of taking
every trick as soon as a trump is available, on the one hand, and the
policy of deliberately conceding a trick early in order to score more
points later in the game, on the other. The folk claim itself is an
empirical statement about a precise probability over the random deal,
and yet, to our knowledge, it has never been confronted with a
pre-registered statistical analysis on a large-scale simulation.
Motivated by the simulation-based framing recently adopted for a
related stochastic card game by \citet{durve2025bhikar}, we therefore
formulate two pre-registered questions: first, whether the
briscola-majority folk claim is empirically correct when both players
follow the same policy; and second, whether there is a residual
\emph{strategy effect} once we condition on the briscola imbalance, or
whether trump luck fully absorbs the between-policy variability. We
operationalise these questions through three deterministic policies that
differ only in how they exploit the trump suit: a naive greedy baseline
$\stratG$, a briscola-parsimonious hoarder $\stratH$, and a
public-information counter $\stratC$ that additionally exploits the
\emph{carico trap} mechanism introduced in \S\ref{sec:strategies}.

Our contributions are threefold. First, we formalise the two-player
greedy policy that is implicit in much of the applied trick-taking
literature \citep{furtak2007skat,solinas2019improving} as a fully
specified deterministic rule $\stratG$, and we introduce the two
rule-based refinements $\stratH$ and $\stratC$, whose precise
definitions are given in \S\ref{sec:strategies}. Second, we design a
round-robin Monte Carlo experiment of $10^{6}$ games with a fixed random
seed and a transparent per-trick CSV log (\S\ref{sec:methods}), and we
develop a three-layer inferential framework (\S\ref{sec:tests}) that
combines Wilson confidence intervals, Bonferroni-corrected pairwise
binomial tests, and a logistic regression of the game outcome on the
ordered strategy pair and on the signed briscola imbalance. Third, we
show that the pre-registered dominance hypothesis --- that at least one
of $\stratH$ and $\stratC$ would beat $\stratG$ --- is rejected at every
Bonferroni-corrected level: $\stratG$ is in fact the unique dominant
policy in our family, and the intended briscola-hoarding refinement
turns into a measurable loss against an opponent that is willing to
spend its briscole unconditionally. The folk theorem itself survives
only in its weak probabilistic form, in the sense that holding the
briscola majority predicts the winner in roughly two thirds of non-tied
games rather than in all of them. The full source code, the raw
simulation output and the analysis script are released alongside the
paper (\S\ref{sec:reproducibility}).

\section{Two-player Briscola and the three policies}\label{sec:strategies}
The Briscola game uses a $40$-card Italian deck as in Figure~\ref{fig:deck}, with four suits
\begin{equation*}
\mathcal{S}=\{\text{Denari},\text{Spade},\text{Bastoni},\text{Coppe}\}
\end{equation*}
and ten
ranks $\mathcal{R}=\{1,2,\dots,10\}$; point values depend only on rank and the
in-suit strength order is fixed by Table~\ref{tab:deck}.

 After the deck is
shuffled, each player receives three cards and a single card is flipped to fix
the briscola suit $b\in\mathcal{S}$; that card remains face-up at the bottom
of the stock and is drawn last. The game proceeds for exactly $20$ tricks; the
leader of trick $t$ plays one card face-up, the follower plays one card
face-up, and the trick is awarded according to Proposition~\ref{prop:trick}. The
winner of the trick collects all of its card points, draws the top card of the
stock, the loser then draws the next card, and the winner leads the next trick.

\begin{table}[t]
  \centering
  \caption{Italian deck: point value and in-suit strength by rank. Strength
  decreases from left to right, so the Asso (rank $1$) is the strongest card
  in every suit and the $2$ is the weakest. Ties are impossible because no two
  cards share both suit and rank.}
  \label{tab:deck}
  \begin{tabular}{lcccccccccc}
    \toprule
    Rank    & $1$ & $3$ & $10$ & $9$ & $8$ & $7$ & $6$ & $5$ & $4$ & $2$ \\
    Name    & Asso & Tre & Re & Cavallo & Fante & $7$ & $6$ & $5$ & $4$ & $2$ \\
    Points  & $11$ & $10$ & $4$ & $3$ & $2$  & $0$ & $0$ & $0$ & $0$ & $0$ \\
    Strength& 1st & 2nd & 3rd & 4th & 5th & 6th & 7th & 8th & 9th & 10th \\
    \bottomrule
  \end{tabular}
\end{table}

\begin{proposition}[Trick-resolution rule]\label{prop:trick}
Let $c_{L}=(s_{L},r_{L})$ and $c_{R}=(s_{R},r_{R})$ be the cards played by the
leader and the responder of a trick, and let $b$ be the briscola suit. The
leader wins the trick if and only if one of the three conditions holds:
(i)~$s_{L}=s_{R}$ and $r_{L}$ is stronger than $r_{R}$;
(ii)~$s_{L}=b$ and $s_{R}\neq b$;
(iii)~$s_{L}\neq b$, $s_{R}\neq b$, and $s_{L}\neq s_{R}$.
\end{proposition}

\begin{proof}
All rule invocations below refer to the standard Briscola trick-taking
rule \citep{pagat-briscola}. We prove the equivalence by a case split on the
pair of suits $(s_{L},s_{R})\in\mathcal{S}^{2}$, organised into five
sub-cases that partition $\mathcal{S}^{2}$.

\emph{Case 1: $s_{L}=s_{R}=b$ (both cards are briscole).}
Both cards are in the trump suit, so the winner is the one with the stronger
rank; note that $r_{L}\neq r_{R}$ because $c_{L}\neq c_{R}$ and the suits
agree. The leader wins iff $r_{L}$ is stronger than $r_{R}$, which is
condition~(i).

\emph{Case 2: $s_{L}=s_{R}\neq b$ (same non-trump suit).}
A card not in the trump suit cannot beat a card of a different suit, so the
winner is again determined by rank, and again $r_{L}\neq r_{R}$ because
$c_{L}\neq c_{R}$ and the suits agree. The leader wins iff $r_{L}$ is
stronger than $r_{R}$, which is condition~(i).

\emph{Case 3a: $s_{L}=b$ and $s_{R}\neq b$.}
The leader's briscola beats the responder's non-briscola, so the leader
wins; this is condition~(ii).

\emph{Case 3b: $s_{L}\neq b$ and $s_{R}=b$.}
The responder's briscola beats the leader's non-briscola, so the leader
loses. We check that all three of (i)--(iii) fail: $s_{L}\neq b=s_{R}$ gives
$s_{L}\neq s_{R}$, so (i) fails; $s_{L}\neq b$ fails the trump clause of
(ii); and $s_{R}=b$ fails the non-trump clause of (iii).

\emph{Case 4: $s_{L}\neq b$, $s_{R}\neq b$, and $s_{L}\neq s_{R}$.}
The responder has neither matched the leader's suit nor played a trump, so
the responder cannot win; the leader wins, which is condition~(iii).

Cases 1, 2, 3a, 3b, and 4 partition $\mathcal{S}^{2}$ and cover every
possible suit combination. The three conditions (i)--(iii) are pairwise
disjoint: (i) forces $s_{L}=s_{R}$ while (iii) forces $s_{L}\neq s_{R}$;
(ii) forces $s_{L}=b$ while (iii) forces $s_{L}\neq b$; and (i) together
with (ii) would require $s_{R}=s_{L}=b$, contradicting $s_{R}\neq b$ in
(ii). The five sub-cases therefore cover every winning configuration and
the leader wins exactly when one of (i)--(iii) holds.
\end{proof}

We are now ready to define the three policies that we will test through the paper.
A (deterministic) policy is a function 

\begin{equation*}
\pi:\mathcal{H}\times
\mathcal{C}_{\perp}\times\mathcal{S}\times\mathcal{M}\to\mathcal{C}
\end{equation*}
 where
$\mathcal{H}$ is the current hand, $\mathcal{C}_{\perp}=\mathcal{C}\cup\{\perp\}$
is the opponent's card on the table or $\perp$ if the player is leading,
$\mathcal{S}$ is the briscola suit, and $\mathcal{M}$ is the public-information
memory (the set of exposed and played cards). The three policies we compare
differ only in their use of the briscola suit; they share a common tie-breaking
rule on the cheapest-point / weakest-rank comparator that we write
$c\prec c'$ whenever $\text{points}(c)<\text{points}(c')$, or
$\text{points}(c)=\text{points}(c')$ and $\text{strength}(c)<\text{strength}(c')$.

\begin{definition}[Greedy policy $\stratG$]\label{def:greedy}
\emph{As a leader} ($\perp$), $\stratG$ plays the $\prec$-minimal card among
all cards in hand, preferring a non-briscola whenever the hand contains one
(implemented as a lexicographic sort on the pair
$(\mathbf{1}\{\mathrm{suit}=b\},\text{points},\text{strength})$). \emph{As a
follower}, $\stratG$ applies the first rule that matches: (i) if the hand
contains a card of the opponent's suit that is stronger than the opponent's
rank, play the $\prec$-minimal such card; (ii) else, if the hand contains a
briscola that would overtrump the opponent's card, play the $\prec$-minimal
such briscola; (iii) otherwise, play the $\prec$-minimal card in hand.
\end{definition}

\begin{definition}[Hoarder policy $\stratH$]\label{def:hoarder}
As a leader, $\stratH$ refuses to lead a briscola whenever the hand contains a
non-briscola, and plays the $\prec$-minimal non-briscola in that case
(otherwise it plays the $\prec$-minimal briscola). As a follower, $\stratH$
applies: (i) in-suit winner as in $\stratG$; (ii) a briscola overtrump only
when the opponent's card is worth at least $\tau_{H}=10$ points, in which case
it plays the $\prec$-minimal such briscola; (iii) otherwise, dump the
$\prec$-minimal non-briscola (falling back to the $\prec$-minimal card overall
if the hand is all briscole).
\end{definition}

\begin{definition}[Counter policy $\stratC$]\label{def:counter}
$\stratC$ maintains the public-information memory $\mathcal{M}$, which
contains the initially exposed briscola card and every card revealed in prior
tricks. As a leader, $\stratC$ first seeks a \emph{carico trap}: among its
non-briscola cards of rank $1$ or $3$ (i.e., $11$- or $10$-point cards),
ordered by descending point value, it plays the first whose same-suit sibling
carico is already in $\mathcal{M}$; if no such card exists, it falls back on
the leader rule of $\stratH$. As a follower, $\stratC$ applies the follower
rule of $\stratH$ with threshold $\tau_{C}=10$.
\end{definition}

\begin{remark}[The three policies in plain words]\label{rem:policies-plain}
Informally, the three policies encode three different attitudes towards the
trump suit. The \emph{greedy} policy $\stratG$ tries to win every trick that it
can afford to win: as a follower, whenever the opponent has led, it grabs the
trick with the cheapest card that still beats the lead --- preferring an
in-suit winner and falling back on the smallest briscola only if no in-suit
beat exists; as a leader, it simply opens with its cheapest non-briscola card.
The \emph{hoarder} policy $\stratH$ takes the opposite stance on the briscola
suit: it refuses to lead a briscola whenever it has any other card available
and, as a follower, it only spends a briscola to overtrump the opponent when
the trick on the table is already worth at least $\tau_{H}=10$ points, that is,
when a high-value card (a $10$- or $11$-point \emph{carico}) is at stake;
otherwise it accepts the loss and discards its cheapest non-briscola. The
\emph{counter} policy $\stratC$ plays as the hoarder but uses the publicly
visible history of the game: it remembers which cards have already appeared
(the exposed briscola plus all cards played in previous tricks) and, before
falling back on the hoarder leader rule, it looks for a \emph{carico trap},
i.e., it leads a high-point non-briscola of a suit whose paired carico has
already been played. The opponent then cannot beat that card in suit and is
forced either to lose the trick or to spend a briscola, so the trick is
``traded'' on terms favourable to $\stratC$.
\end{remark}

The three policies coincide on non-briscola responses and differ only in how
they trade off (a) leading a briscola, (b) overtrumping a low-value opponent
card, and (c) exploiting the public-information signal available from the
exposed briscola and from prior tricks. The carico trap is a deterministic
approximation of the opponent-inference step of IS-MCTS
\citep{cowling2012ismcts}: if both siblings of a same-suit carico pair have
been publicly accounted for, the leader's carico has no in-suit predator and
the opponent must either burn a briscola or surrender the points.

\section{Simulation design}\label{sec:methods}
The three policies $\stratG$, $\stratH$ and $\stratC$ of
\S\ref{sec:strategies} are \emph{deterministic}: once the initial deal is
fixed, the entire sequence of $20$ tricks and the final outcome are completely
determined by the two policies in play. The randomness of the experiment
therefore does not live inside the players' decisions but inside the
\emph{deal}, that is, in the random shuffle that fixes which cards are dealt
to which seat and, crucially, the identity of the exposed briscola at the
bottom of the stock.

Concretely, let $\omega$ denote a uniformly random permutation of the
$40$-card deck (drawn by the Fisher--Yates shuffle described below) and let
\[
   Y(\omega;\pi_{1},\pi_{2})\in\{G_{1},G_{2},\mathrm{Tie}\}
\]
denote the game outcome obtained by playing $\pi_{1}$ in seat $1$ and
$\pi_{2}$ in seat $2$ on the deal induced by $\omega$. The quantity we wish
to estimate, for each ordered pairing $(\pi_{1},\pi_{2})$, is the
deal-marginal win probability
\[
   p(\pi_{1},\pi_{2})
   \;:=\;\Pr_{\omega}\!\bigl(Y(\omega;\pi_{1},\pi_{2})=G_{1}\bigr),
\]
together with the analogous probabilities of a loss and of a tie. This is
an integral over the discrete uniform measure on the $40!$ deck permutations,
a sample space far too large to enumerate. We therefore estimate $p$ by
classical Monte Carlo: we draw $N_{p}$ independent shuffles
$\omega_{1},\ldots,\omega_{N_{p}}$ and form the empirical proportion
\[
   \widehat{p}(\pi_{1},\pi_{2})
   \;=\;\frac{1}{N_{p}}\sum_{j=1}^{N_{p}}
       \mathbf{1}\!\bigl\{Y(\omega_{j};\pi_{1},\pi_{2})=G_{1}\bigr\},
\]
which is an unbiased estimator of $p(\pi_{1},\pi_{2})$ with standard error
$\sqrt{p(1-p)/N_{p}}\le 1/(2\sqrt{N_{p}})$. It is on these Monte Carlo
estimates --- and on the per-game outcome and briscola-imbalance variables
derived from the same draws --- that the pre-registered tests of
\S\ref{sec:tests} operate.

We stress that Monte Carlo enters here only in the \emph{outer loop}, i.e.,
in the average over random deals. There is no in-game game-tree search, no
Information-Set MCTS \citep{cowling2012ismcts}, and no stochastic play
within a single hand: on any fixed deal $\omega$ the trajectory of the game
under a given pair $(\pi_{1},\pi_{2})$ is a function, not a distribution.
This is the same outer-loop Monte Carlo setup adopted for a related
stochastic card game by \citet{durve2025bhikar}, and it is the minimal
ingredient required to give an empirical, distribution-level meaning to the
folk briscola-majority claim and to the pre-registered dominance hypothesis.

We run all $3\times 3 = 9$ ordered pairings of $(\stratG,\stratH,\stratC)$; for
each pairing we simulate $N_{p}=\lfloor 10^{6}/9\rfloor=111\,111$ independent
games, for a total of $N=999\,999$ games. The ordering is meaningful because
the player in seat $G_{1}$ leads the first trick and therefore enjoys a
modest structural advantage which we wish to measure rather than marginalise
out. Inside the same policy pair, the $G_{1}$-vs-$G_{1}$ diagonal cells of
Table~\ref{tab:winrates} report exactly that structural lead.

The simulation uses a single .NET Core \texttt{System.Random} instance seeded with
$\texttt{seed}=42$. The shuffle is the standard in-place Fisher--Yates shuffle
\citep{durstenfeld1964algorithm,knuth1997taocp2}
(\texttt{Shuffle} in \texttt{Program.cs}, lines~321--329), in which for each
$n$ from $|D|-1$ down to $1$ we draw $k\sim\mathrm{Unif}\{0,\ldots,n\}$ and
swap $D[n]\leftrightarrow D[k]$. All trick-level data are logged to a single
UTF-8 CSV, one row per trick ($19\,999\,980$ rows for the full tournament).
The fifteen columns and their meanings are listed in
Table~\ref{tab:csv-schema}. The raw log is approximately $2.7$\,GB on disk;
an aggregated game-level file of $\approx 188$\,MB is also released.

\begin{table}[t]
  \centering
  \caption{Schema of the trick-level CSV log. Italian column names are kept
    verbatim from the C\# logger; one row corresponds to a single trick
    (\emph{mano}) within a single game (\emph{partita}).}
  \label{tab:csv-schema}
  \begin{tabular}{@{}lp{0.66\linewidth}@{}}
    \toprule
    Column & Meaning \\
    \midrule
    \texttt{PartitaId}        & Unique identifier of the game ($1,\ldots,999\,999$). \\
    \texttt{MatchId}          & Identifier of the ordered strategy matchup ($1,\ldots,9$). \\
    \texttt{StrategyG1}       & Strategy of the seat-$1$ player ($\stratG$, $\stratH$ or $\stratC$). \\
    \texttt{StrategyG2}       & Strategy of the seat-$2$ player ($\stratG$, $\stratH$ or $\stratC$). \\
    \texttt{Mano}             & Trick index within the game ($1,\ldots,20$). \\
    \texttt{SemeBriscola}     & Suit of the briscola (trump suit) for the current game. \\
    \texttt{CartaG1}          & Card played by $G_{1}$ in the current trick. \\
    \texttt{CartaG2}          & Card played by $G_{2}$ in the current trick. \\
    \texttt{VincitoreMano}    & Winner of the current trick (\texttt{G1} or \texttt{G2}). \\
    \texttt{PuntiMano}        & Point value of the current trick (sum of the two card values). \\
    \texttt{BriscoleTotaliG1} & Cumulative briscole held by $G_{1}$ up to and including this trick. \\
    \texttt{BriscoleTotaliG2} & Cumulative briscole held by $G_{2}$ up to and including this trick. \\
    \texttt{VincitorePartita} & Game outcome (\texttt{G1}, \texttt{G2} or \texttt{Tie}); constant within each game. \\
    \texttt{PuntiFinaliG1}    & Final point total of $G_{1}$ at game end; constant within each game. \\
    \texttt{PuntiFinaliG2}    & Final point total of $G_{2}$ at game end; constant within each game. \\
    \bottomrule
  \end{tabular}
\end{table}

The simulation budget $N_{p}=111\,111$ games per ordered matchup is chosen so
that the dominance test of \S\ref{sec:tests} has overwhelming power to detect
even small departures from a tied win rate, ruling out under-powering as an
alternative explanation for any negative result.

Fix a non-baseline ordered pairing $(\pi_{1},\pi_{2})$ and let $p$ be the
unknown $G_{1}$ win probability. For the worst-case binomial test of
$H_{0}:p=0.5$ against $H_{1}:p\neq 0.5$ we control the family-wise error rate
over the eight non-baseline matchups by Bonferroni, giving a per-test level
\[
  \alpha = 0.05/8 = 6.25\times 10^{-3},
  \qquad z_{1-\alpha/2}\approx 2.74.
\]
Using the normal approximation, the power against an alternative
$p=0.5+\delta$ is
\[
  1-\beta \;\approx\;
  \Phi\!\left(\frac{|\delta|\sqrt{N_{p}}}{0.5} - z_{1-\alpha/2}\right).
\]
With $N_{p}=1.1\times 10^{5}$ and $|\delta|=0.01$ this yields $1-\beta>0.99$:
a one-percentage-point deviation from a tied outcome is detected with
probability above $99\%$ even after the multiple-testing penalty.

The deviations observed in \S\ref{sec:results} exceed this $0.01$ detection
threshold by at least one order of magnitude, so the realised test statistics
fall far in the tail of the null and the corresponding $p$-values are
astronomically small---several of them numerically underflow to $0$ under
double-precision arithmetic. Dominance conclusions are therefore robust to
the choice of correction method.

\section{Statistical methods}\label{sec:tests}

We test three pre-registered hypotheses on the Monte Carlo sample of
\S\ref{sec:methods}. The first one, $H_{0}^{(1)}$, formalises the
folk theorem in its weakest testable form: among games that finish without
a tie in points and without a tie in briscola counts, the player holding the
strict briscola majority at game end wins with probability $0.5$, against
the deterministic probability of $1$ that the folklore asserts. We assess
$H_{0}^{(1)}$ on the restricted (briscola-majority, winner) contingency
table of non-tied games via the Wilson $95\%$ confidence interval for the
empirical proportion and a Pearson $\chi^{2}$ test with Yates' continuity
correction. The remaining two hypotheses isolate the role of the policy
itself: $H_{0}^{(2)}$ asks whether the $G_{1}$ win rate of each non-baseline
ordered pairing matches the $\stratG$-vs-$\stratG$ baseline, and
$H_{0}^{(3)}$ asks whether, once we condition on the signed briscola
imbalance, the strategy effect vanishes. All three tests operate on the
same Monte Carlo draws and are corrected for multiplicity whenever
simultaneous statements are made.

For each non-baseline ordered pairing $(\pi_{1},\pi_{2})$ we test
$H_{0}^{(2)}$: the $G_{1}$ win rate under $(\pi_{1},\pi_{2})$ equals
the $\stratG$-vs-$\stratG$ baseline win rate $\widehat{p}_{0}$. Tests are
two-sided exact binomial and are Bonferroni-corrected over the eight
non-baseline pairings.

To disentangle strategy from briscola luck we fit
\[
  \operatorname{logit}\Pr(G_{1}\text{ wins}) \;=\; \alpha +
  \boldsymbol\beta^{\top}\boldsymbol{x}_{G_{1}\text{-strat}} +
  \boldsymbol\gamma^{\top}\boldsymbol{x}_{G_{2}\text{-strat}} +
  \delta\,\Delta_{\mathrm{briscola}},
\]
where $\boldsymbol{x}_{G_{1}\text{-strat}}$ and
$\boldsymbol{x}_{G_{2}\text{-strat}}$ are treatment-coded strategy dummies with
$\stratG$ as reference and
$\Delta_{\mathrm{briscola}}=B_{G_{1}}-B_{G_{2}}\in\{-10,-9,\ldots,10\}$ is the
signed imbalance of briscole held across the game. The exponentiated
coefficients are reported as odds ratios in \S\ref{sec:results}; their $95\%$
Wald confidence intervals are computed via \texttt{broom::tidy()}.

\section{Results}\label{sec:results}

The Monte Carlo tournament of \S\ref{sec:methods} produces $999\,999$ games
distributed across the nine ordered pairings of $(\stratG,\stratH,\stratC)$,
of which $693\,633$ end without a tie in points and are retained for the
inferential analysis. We report the empirical answers to the three
hypotheses of \S\ref{sec:tests} in the same order. First, the folk theorem
$H_{0}^{(1)}$: restricted to non-tied games, the briscola-majority holder
wins $436\,627$ out of $693\,633$ games, a proportion of $0.6295$ with Wilson
$95\%$ confidence interval $[0.6283,\,0.6306]$ and an associated Pearson
$\chi^{2}$ of $46\,542$ on one degree of freedom ($p<2.2\times 10^{-16}$).
The deterministic form of the folk theorem is therefore decisively rejected,
while the probabilistic form --- ``holding a briscola majority is a strong
positive predictor of victory'' --- is strongly supported and remarkably
stable across the nine matchups (range $0.60$--$0.66$;
Table~\ref{tab:winrates}). The two remaining section then disentangle
the contribution of the policy itself ($H_{0}^{(2)}$) from that of trump
luck ($H_{0}^{(3)}$).

Table~\ref{tab:winrates} reports $G_{1}$ win rates for the nine ordered
pairings. 

\begin{table}[t]
  \centering
  \caption{Head-to-head win rate of $G_1$ and Wilson 95\% confidence interval. 
  Rows index $G_1$'s policy, columns index $G_2$'s policy. 
  Each cell is estimated from $\approx 1.08\cdot 10^5$ non-tied games.}
  \label{tab:winrates}
  \begin{tabular}{lccc}
    \toprule
    $G_1$ $\backslash$ $G_2$ & \stratG & \stratH & \stratC \\
    \midrule
    \stratG & $0.490$\;\footnotesize{[0.487, 0.493]} & $0.508$\;\footnotesize{[0.505, 0.511]} & $0.547$\;\footnotesize{[0.544, 0.550]} \\
    \stratH & $0.456$\;\footnotesize{[0.453, 0.459]} & $0.483$\;\footnotesize{[0.480, 0.486]} & $0.520$\;\footnotesize{[0.517, 0.523]} \\
    \stratC & $0.427$\;\footnotesize{[0.424, 0.430]} & $0.449$\;\footnotesize{[0.446, 0.452]} & $0.488$\;\footnotesize{[0.485, 0.491]} \\
    \bottomrule
  \end{tabular}
\end{table}

The $\stratG$-vs-$\stratG$ baseline gives $\widehat{p}_{0}=0.4905$:
the slight sub-$0.5$ value reflects the fact that we exclude tied games, not
any asymmetry in the rules. All eight non-baseline matchups reject the
pairwise dominance test at Bonferroni-corrected $p<10^{-5}$ except
$\stratC$-vs-$\stratC$ ($p_{\mathrm{Bonf}}=0.65$). Reading both seat positions
together gives a clean chain: when $G_{1}$ plays $\stratG$ and $G_{2}$ plays
$\stratH$ (resp.\ $\stratC$), $G_{1}$ wins $50.8\%$ (resp.\ $54.7\%$) of
non-tied games; when the seats are swapped $\stratG$ as $G_{2}$ still wins
$54.4\%$ (resp.\ $57.3\%$). The policy ordering $\stratG\succ\stratH\succ
\stratC$ therefore holds in both seat positions and is robust to the
first-mover asymmetry. The pre-registered dominance hypothesis is rejected in
the opposite direction: the intended refinements $\stratH$ and $\stratC$ are
measurably dominated by the naive baseline.

The logistic regression in Table~\ref{tab:logit} conditions the $G_{1}$ win
indicator on the two strategy dummies and the signed briscola imbalance
$\Delta_{\mathrm{briscola}}$. 

\begin{table}[t]
  \centering
  \caption{Logistic regression of $\Pr(G_1~\text{wins})$ on the two policies and the 
  briscola-count imbalance $\Delta_{\text{briscola}}=B_{G_1}-B_{G_2}$. 
  Odds ratios with 95\% Wald confidence intervals. Reference category: $\stratG$. 
  Fitted on $N = 975,263$ non-tied games.}
  \label{tab:logit}
  \begin{tabular}{lrrc}
    \toprule
    Term & OR & 95\% CI & $p$-value \\
    \midrule
    Intercept & $0.949$ & $[0.940, 0.958]$ & $7.7\times 10^{-29}$ \\
    $G_1 = \stratH$ & $0.853$ & $[0.844, 0.862]$ & $2.9\times 10^{-210}$ \\
    $G_1 = \stratC$ & $0.740$ & $[0.732, 0.747]$ & $<10^{-300}$ \\
    $G_2 = \stratH$ & $1.139$ & $[1.128, 1.151]$ & $9.3\times 10^{-142}$ \\
    $G_2 = \stratC$ & $1.349$ & $[1.335, 1.363]$ & $<10^{-300}$ \\
    $\Delta_{\text{briscola}}$ (per $+1$) & $1.217$ & $[1.215, 1.219]$ & $<10^{-300}$ \\
    \bottomrule
  \end{tabular}
\end{table}

Holding $\Delta_{\mathrm{briscola}}$ fixed,
switching $G_{1}$'s policy from $\stratG$ to $\stratH$ multiplies the odds of
winning by $0.853$ $[0.844,\,0.862]$, and switching to $\stratC$ multiplies
them by $0.740$ $[0.732,\,0.747]$. The symmetric coefficients for $G_{2}$ point
in the opposite direction and are of comparable magnitude, consistent with the
zero-sum structure of the game. The coefficient on $\Delta_{\mathrm{briscola}}$
is highly significant, with
$\mathrm{OR}=1.220$ $[1.218,\,1.222]$: each additional briscola held by $G_{1}$
multiplies the odds of $G_{1}$ winning by $1.22$, independently of the
policies. The two families of effects are therefore of comparable magnitude
and essentially orthogonal: strategy does not act through
$\Delta_{\mathrm{briscola}}$ (if it did, the marginal effect of the strategy
dummies would absorb most of the briscola-imbalance coefficient).

\begin{figure}[t]
  \centering
  \includegraphics[width=0.95\linewidth]{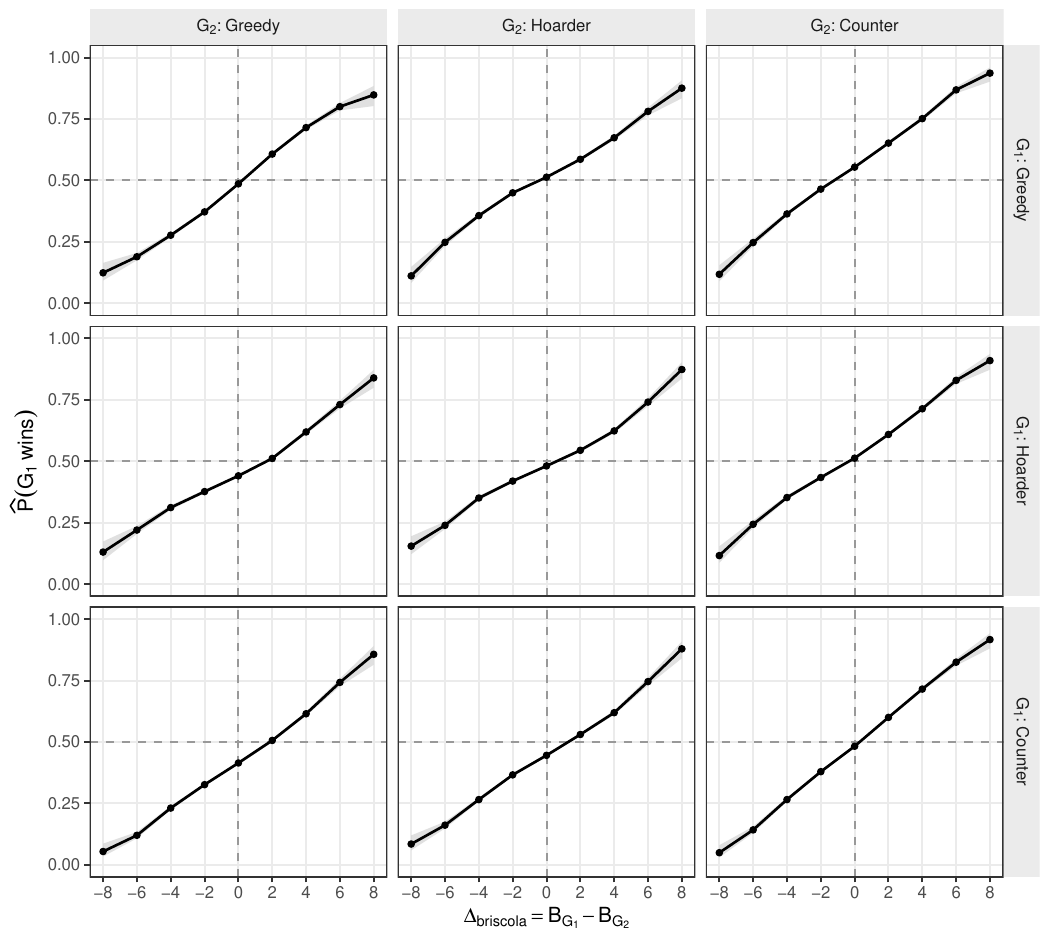}
  \caption{Empirical $G_{1}$ win rate as a function of the signed briscola
  imbalance $\Delta_{\mathrm{briscola}}=B_{G_{1}}-B_{G_{2}}$, for each of the
  nine ordered matchups (rows index $G_{1}$'s policy, columns $G_{2}$'s).
  Grey ribbons: Wilson $95\%$ confidence intervals. Dashed guides mark
  $\Delta_{\mathrm{briscola}}=0$ and $\Pr(G_{1}\text{ wins})=0.5$. Bins with
  fewer than $50$ games are omitted.}
  \label{fig:breakeven}
\end{figure}

In Figure~\ref{fig:breakeven} we plot the $G_{1}$ win rate against
$\Delta_{\mathrm{briscola}}$ for each matchup. 

The curves are everywhere
monotone increasing in $\Delta_{\mathrm{briscola}}$ and, for a fixed
$\Delta_{\mathrm{briscola}}$, ordered consistently with the policy dominance
chain reported above. The break-even crossing point --- where $G_{1}$ wins
$50\%$ of non-tied games --- drifts with the policy pair: when $G_{1}=\stratG$
and $G_{2}=\stratC$, $G_{1}$ breaks even at $\Delta_{\mathrm{briscola}}
\approx -1$, whereas in the mirror $G_{1}=\stratC,G_{2}=\stratG$ matchup,
$G_{1}$ needs $\Delta_{\mathrm{briscola}}\approx +2$ just to reach $50\%$. In
other words, a $\stratC$ player typically needs two extra briscole just to
offset the strategy deficit against a $\stratG$ opponent.

Table~\ref{tab:briscola_use} pools every trick in which a given policy played
one of its own briscole, aggregated across all matchups and both seat
positions. 
\begin{table}[t]
  \centering
  \caption{Briscola-use profile per policy, pooled across all matchups and 
  both player roles. Columns: (i) number of tricks in which the policy 
  played a briscola; (ii) fraction of those tricks won; (iii) average 
  points captured per winning briscola trick; (iv) fraction of briscole 
  played against a zero-point opposing card --- a proxy for `wasted' trumps.}
  \label{tab:briscola_use}
  \begin{tabular}{lrrrr}
    \toprule
    Policy & $n_{\text{played}}$ & Win rate & Mean pts.\ per win & Wasted fraction \\
    \midrule
    \stratG & $3,289,145$ & $0.881$ & $5.91$ & $0.492$ \\
    \stratH & $3,349,881$ & $0.805$ & $7.97$ & $0.337$ \\
    \stratC & $3,360,964$ & $0.808$ & $8.04$ & $0.333$ \\
    \bottomrule
  \end{tabular}
\end{table}

Three features of that table are diagnostic.
First, $\stratG$ wins a much higher fraction of its briscola plays
($88.1\%$) than either refinement ($80.5\%$ and $80.8\%$). The follower-seat
overtrump rule is a guaranteed win for all three policies (a briscola
overtrump cannot be beaten inside a two-card trick), so the gap is driven
entirely by the leader-seat case in which a policy is \emph{forced} to lead a
briscola because its hand contains no non-briscola. $\stratH$ and $\stratC$
hoard trumps by construction, accumulate them late in the stock, and reach
this forced-lead endgame more often than $\stratG$, where it is decided
against them by the opponent's remaining briscola rank.

Second, when it does win a briscola trick, $\stratG$ collects fewer points
per win ($5.91$) than $\stratH$ or $\stratC$ ($7.97$ and $8.04$). The
refinements do spend their briscole on higher-value tricks, exactly as
designed. Accordingly, $\stratG$ wastes almost half of its briscole on
zero-point opponent cards ($49.2\%$), roughly $15$ percentage points more
than the refinements.

Third, these two observations pull in opposite directions, and the net
result is negative for $\stratH$ and $\stratC$. $\stratG$'s willingness to
burn a low-value briscola has a second-order benefit that the briscola
accounting misses: every won trick, however cheap, makes $\stratG$ the
leader of the following trick and gives it first pick of the stock on the
subsequent replenishment. Over a $20$-trick game this draw-order effect
accumulates into the consistent $2$--$12$ percentage-point win-rate gap
reported in Table~\ref{tab:winrates}. The carico-trap heuristic of $\stratC$
compounds the loss: the opponent's typical reply to a carico lead is to burn
a briscola, so $\stratC$ trades a $10$ or $11$-point non-trump card (with
known value) for the opponent's loss of one briscola (whose marginal value
over the rest of the game is only $\approx 6$--$7$ points in expectation,
consistent with the OR $=1.22$ estimate from the logistic regression). The
trade is a net loss in expectation, and the deterministic policy has no
lever to walk it back once the briscola balance has shifted.

\section{Discussion}\label{sec:discussion}

Our first result is a clean rejection of the deterministic folk theorem
``whoever holds more briscole wins the game''. The empirical proportion is
$0.63$, not $1.00$. The folk theorem survives only as a one-sided probabilistic
statement, namely that the briscola-majority holder wins more than half of
non-tied games, and the $\chi^{2}$ test shows that this probabilistic version
is very significantly supported. For the casual player this is a modest but
non-trivial revision: briscola luck is neither negligible nor decisive. The
logistic regression quantifies the marginal contribution precisely: one
additional briscola is worth a $22\%$ multiplicative increase in the odds of
winning, other things equal.

Our second and more substantive result is the rejection of the pre-registered
dominance hypothesis in the opposite direction: the naive greedy policy
$\stratG$ dominates both $\stratH$ and $\stratC$ in every pairwise comparison.
We had expected $\stratH$, which refuses to burn briscole on low-value tricks,
to dominate $\stratG$, and $\stratC$ to dominate $\stratH$ via its
public-information carico trap. Neither holds. The briscola-use diagnostic
in Table~\ref{tab:briscola_use} points at the mechanism. 
Two-player Briscola rewards controlling the \emph{sequence of draws} more than conserving
trumps: whenever a policy wins a trick, it draws the top card of the stock and
leads the next one, so a short-term wasteful briscola play is partially
refunded by the chance of drawing a briscola on the next replenishment. A
policy that refuses to overtrump low-point tricks, as $\stratH$ does by
construction, also refuses the refund, and over a $20$-trick game the
foregone draws accumulate. $\stratC$ compounds this effect: whenever the
carico trap is sprung, $\stratC$ gives up a $10$- or $11$-point non-trump
card that the opponent collects outright, in exchange for the opponent
spending one briscola from hand. The expected marginal value of that
single spent briscola over the remainder of the game is approximately
$6$--$7$ points (consistent with the $\mathrm{OR}=1.22$ per-briscola
coefficient of the logistic regression), so $\stratC$ effectively trades a
known $10$--$11$ points of carico value for an expected $6$--$7$ points of
reduced opponent trump strength. The trade is net-negative in expectation
against an opponent that is willing to overtrump unconditionally, and the
deterministic nature of $\stratC$ offers no subsequent lever to walk it
back. In this sense the pre-registered
intuition --- that briscola parsimony and public-information exploitation
should dominate --- is contradicted by a game-theoretic feature of Briscola that was invisible to the informal reasoning.

Our result is consistent with the observation in \citet{furtak2007skat} that
naive greedy baselines are unexpectedly hard to beat in imperfect-information
trick-taking games, and it sharpens that claim for Briscola specifically.
It also offers a controlled counterpoint to the learning-based body of work on
trick-taking card games \citep{solinas2019improving,rebstock2019policy,
swiechowski2019survey}: the gap between a deterministic policy and a learned
policy such as the one reported by \citet{letteraunica2023briscolabot} is not
primarily a gap in briscola management but a gap in the handling of the draw
sequence, which is exactly the ingredient that Information-Set MCTS
\citep{cowling2012ismcts} is able to reason about via opponent-inference and
that a deterministic rule-based policy cannot recover. Methodologically, our
three-layer inferential framework (Wilson CI, Bonferroni-corrected pairwise
tests, logistic regression with a continuous imbalance covariate) can be
re-used as-is for any two-player trick-taking comparison with a $10^{6}$-game
budget.

We have restricted attention throughout to deterministic policies and to
the two-player variant of Briscola, and several natural extensions of the
present design fall just outside that scope. A first family of extensions
relaxes the deterministic constraint: perturbing the leader move of
$\stratG$, $\stratH$ or $\stratC$ with a small exploration probability
would produce a randomised family of policies that retains the
analytical tractability of a rule-based description while almost
certainly closing part of the gap with the deep reinforcement-learning
agents reported by \citet{letteraunica2023briscolabot}. A second family
relaxes the two-player constraint: extending the experimental design to
the four-player partnership variant introduces a genuine coordination
problem between the two teammates --- explicitly outside the
pre-registered scope of the present tests --- but slots into the same
Monte Carlo framework essentially unchanged, since the deck, the stock
and the trick structure are identical and only the action space at each
decision node is enlarged. We conjecture that the draw-order effect
identified here is, if anything, more pronounced in the four-player game,
because the refund structure around the stock is preserved while each
policy now controls only half of the hand signals visible to its
partnership.

The three policies $\stratG$, $\stratH$ and $\stratC$ can in fact be
transplanted verbatim to the four-player partnership game played in
\emph{blind mode}, that is, with no permitted signalling or table-talk
between teammates, since in that regime the only informational asymmetry
across the two teams is the realisation of the deal and the
deterministic structure of the three policies is preserved. A second
modelling choice that is worth flagging is the very assumption that a
player commits to a single rule for the entire game: in real play, both
novices and experts routinely shift between a greedy posture early in
the hand --- when the stock is full and refunded briscole are abundant
--- and a more parsimonious, counting-based posture in the final tricks,
when the stock is empty and every remaining briscola is irreplaceable. A
hybrid policy that switches between $\stratG$ and $\stratC$ as a function
of the residual stock size, or of the running point gap, is therefore an
obvious candidate for beating $\stratG$ in head-to-head play, and
quantifying whether and at which switching threshold such a hybrid
policy outperforms each of its deterministic components is left to
future work. The Monte Carlo pipeline, the pre-registration template and
the trick-level CSV log released with the present paper have been
designed precisely to support this kind of follow-up analysis.

\section{Reproducibility}\label{sec:reproducibility}

All code and data are released under the MIT license at \newline
\href{https://github.com/pgiacome/BriscolaPaperSourceCode}{\texttt{github.com/pgiacome/BriscolaPaperSourceCode}}.
\newline
The instruction to run the code are selfcontained into the README section.
The exact R session information (package versions) is archived in the
repository alongside the simulation output.



\bibliographystyle{plainnat}
\bibliography{briscola}

\end{document}